\documentclass[12pt]{iopart}
\usepackage{bm,graphicx,amssymb,subfig,psfrag}

\newcommand{\sG}{\mathcal{G}}
\newcommand{\bx}{\bm{x}}
\newcommand{\by}{\bm{y}}
\newcommand{\tP}{\widetilde{P}}

\begin{document}
\title{Maximum-entropy moment-closure for stochastic systems on networks}
\author{Tim Rogers}
\address{School of Physics \& Astronomy, The University of Manchester,
M13 9PL, UK}
\ead{tim.rogers@manchester.ac.uk}
\begin{abstract}
Moment-closure methods are popular tools to simplify the mathematical analysis of stochastic models defined on networks, in which high dimensional joint distributions are approximated (often by some heuristic argument) as functions of lower dimensional distributions. Whilst undoubtedly useful, several such methods suffer from issues of non-uniqueness and inconsistency. These problems are solved by an approach based on the maximisation of entropy, which is motivated, derived and implemented in this article. A series of numerical experiments are also presented, detailing the application of the method to the Susceptible-Infective-Recovered model of epidemics, as well as cautionary examples showing the sensitivity of moment-closure techniques in general.

\end{abstract}
\submitto{JStat}

\section{Introduction}
A great many problems of interest in statistical physics and applied mathematics may be modelled as a system of stochastic variables whose interactions form a network. The earliest such models were inspired by questions in chemistry and typically imposed a lattice interaction structure (see \cite{Ziman1979} and references therein for many examples), though as the applications of these models have diversified to include everything from finance \cite{May2008} to ecology \cite{Dieckmann2000}, the range of network topologies under consideration has expanded also. Whilst of great benefit to the real-world relevance of networked models, the complexity of the network structures now in vogue causes serious complications to any attempted mathematical analysis.\par
In the absence of a network structure, the behaviour of well-mixed stochastic models may often be very closely described by systems of differential equations, which have proven to be indispensable tools in their analysis. It is a long-standing problem to reproduce this success in the field of networked models. The primary obstacle to this endeavor is the existence of correlations between spatially separated parts of the networked model. In mathematical terms, one finds that differential equations describing the joint distribution of some number of variables will typically depend on joint distributions of higher dimension, leading to a never ending expansion in the number of equations required \footnote{Alternatively, consideration of the time evolution of macroscopic observables in disordered systems on networks leads to a subtly different closure problem which has been addressed in, e.g. \cite{Mozeika2008,Mozeika2009}.}.\par
For as long as this problem has been recognised, there have been approximation schemes suggested to deal with it. Often trading under the name `moment-closure' these techniques amount to choosing a method by which to guess at higher dimensional distributions, based on knowledge of the lower dimensional distributions for which one wishes to write differential equations. \par
The simplest non-trivial example of a moment-closure scheme is the pair approximation, proposed for ecological applications in  \cite{Matsuda1992,Satulovsky1994,Keeling1997}, but already with a long history in statistical physics (see, for example \cite{Kikuchi1951,Dickman1986,Tome1989}). In this approximation, the state of the whole system is reduced to the joint probability of pairs of variables which neighbour each other in the network. The joint distribution of three connected variables is then approximated by one of two particular functions, depending on whether they form a triangle or a line in the network. This approximation has been widely applied and is successful for a reasonably broad range of models and parameters \cite{Dieckmann2000,Levin1996,DeAguiar2003,Rozhnova2010}, particularly in those of adaptive networks \cite{Gross2006,Gross2008,Zschaler2010,Demirel2011}. There is, however, a mathematical problem with the pair approximation as it is usually applied to triangles. The formula given in \cite{Keeling1997,Matsuda2000} for the joint distribution of three variables forming a triangle is in general not a probability distribution at all (it is not normalised) and is typically inconsistent with the known marginal distribution of pairs. We will address this issue in detail in Sections 2.4 and 3.2. \par
Moving beyond the pair approximation, more complex schemes have been proposed based on joint probabilities of three or more variables; examples include those of references \cite{Rozhnova2010,Szabo2004,Petermann2004}. The moment-closure methods employed in most such works are reminiscent of Bayes nets \cite{Pearl1988}, being based on the assumption of some form of conditional independence, allowing the factorisation of high dimensional joint probabilities into smaller objects. Therein lies a problem of non-uniqueness: depending on the local configuration of the network, there may be several such factorisations to choose from and no logical motivation for making one choice over another. For a discussion of this problem in relation to an approximation based on triples, see \cite{Rozhnova2010}. \par
This article will develop a moment-closure approach based on the maximisation of entropy. Put simply, when the need arises to approximate a high dimensional distribution based on knowledge of some lower dimensional marginals, it is wise to avoid introducing any additional, unmotivated, bias; this is achieved by choosing the distribution of maximum entropy subject to the known constraints. The approach has the distinct advantages of providing a unique and consistent approximation scheme for distributions of any dimension, and numerical experiments demonstrate considerable success in certain models. There is, however, the aesthetic downside that often the resulting approximation cannot be written as an explicit function of the lower dimensional distributions. In these circumstances, one must deal with it either implicitly or numerically; some appropriate methods are outlined in Section 2.3. It should be noted that the idea of applying maximum-entropy to obtain moment-closure approximations is not unique to this article, having been discussed recently for continuous space models in physics \cite{Singer2004} and ecology  \cite{Raghib2010}. The emphasis here is on the application of the method to networked models, and the relationship to existing alternative schemes.\par
To connect with other approaches, the maximum-entropy formalism is applied to the pair approximation in Section 2.4. In dealing with the case of three sites in a row the pair and maximum-entropy approximations agree, whilst for triangles we are able to suggest an alternative to the inconsistent choice usually made. The performance of this proposed alternative approximation is checked numerically in Section 3.2, with the surprising result that, whilst it is considerably better at predicting the joint distribution on triangles, the output from numerical integration of the moment-closed differential equations is notably worse. Further numerical investigation reveals that this phenomenon can be attributed to a bizarrely fortunate cancellation of error which occurs within the usual pair approximation when applied to networks with many triangles. \par
To bring to a close the numerical investigations of the article, an experiment is proposed to test the robustness of moment-closure methods in general. The basic philosophy of moment-closure rests on the assumption that having a good approximation to the required higher dimensional joint distributions will endow the moment-closed system of differential equations with high predictive power. Despite its central importance to the method, the nature of the relationship between accuracy of approximation and predictive success has not been well studied. In Section 3.3 the results of a simple experiment are presented in which higher dimensional distributions are not approximated but rather measured from a simulation. It is shown that the application of a small systematic error in one direction or another induces a rather large discrepancy in the output of numerical integration of the moment-closed differential equations. This result suggests that the quality of predictions made by moment-closure approximations may be considerably more fluid than one might hope.

\section{Maximum-entropy approximation}
\subsection{Description of the problem}
We consider a discrete time stochastic process \footnote{No generality is lost here by not discussing continuous time processes as both formulations have the same first order behaviour under system size expansion.} defined on a simple graph $\sG=(V,E)$. For notational simplicity, we label the sites by natural numbers: $V=\{1,\ldots,N\}$, where $N$ is the size of the graph. If $(i,j)\in E$ we say that $i$ and $j$ are neighbours in network, and we write $\partial i$ for the set of all neighbours of $i$. Upon each site $i$ a time dependent variable $X_i$ is placed, taking values in a finite set of states. In each timestep, a site $i$ is chosen at random and the state of the variable $X_i$ is updated. The new value is taken to depend only on the old value and those of its neighbours. \par
A great number of processes of interest in statistical physics may be modelled according to this broad description. To fix ideas, we will refer throughout this article to the paradigmatic example of the Susceptible-Infected-Recovered (SIR) model of epidemics. Some number $N$ of individuals are arranged in a network, with the presence of an edge indicating the possibility of infectious contact. The possible states of an individual are denoted $S$, $I$ and $R$, for susceptible, infected and recovered, respectively. Per timestep a site $i$ is chosen at random and the variable $X_i$ updated according to: if $X_i=I$ then we set $X_i=R$ with probability $\gamma$; if $X_i=R$ then we set $X_i=S$ with probability $\alpha$; if $X_i=S$ then a neighbour $j\in\partial i$ is chosen at random and if $X_j=I$ then we set $X_i=I$ with probability $\beta$. The last reaction of this list models the transmission of the disease, which is restricted to pairs of neighbouring sites and thus the spread of the disease may be strongly dependent on the geometry of the network. \par
Rather than always working with the detailed information of the exact state of each variable in a stochastic process, it is usual to consider macroscopic observables -- statistical measures which summarise some aspect of the state of the system at a given time. The simplest example is the distribution of states of a randomly chosen site, we write
\begin{equation*}
P(x)=\frac{1}{|V|}\sum_{i\in V} \mathbb{I}\Big\{X_i=x\Big\}\,,
\end{equation*}
where $\mathbb{I}$ is the indicator function. Objects of this type are very useful in giving an overview of the behaviour of the system, for example, the progress of an SIR epidemic may be monitored by plotting the time evolution of $P(I)$, that is, the fraction of the population currently infected.\par
To gain more detailed information about the underlying process, the above construction may be generalised to arbitrary arrangements of variables in the network. Specifically, for a given (assumed small) graph $g$, let $P_g(\bx)$ denote the fraction of induced subgraphs $h\subset \sG$ which are isomorphic to $g$ and whose site variables take exactly the states $\bx=(x_1,x_2,\ldots)\,$. Formally, we write
\begin{equation}
P_g(\bx)=\frac{\displaystyle\sum_{h\subset \sG}\mathbb{I}\Big\{h\cong g\Big\}\prod_{i\in h}\mathbb{I}\Big\{X_i=x_i\Big\} }{\displaystyle\sum_{h\subset \sG}\mathbb{I}\Big\{h\cong g\Big\}}\,,
\label{emp}
\end{equation}
where $\cong$ denotes graph isomorphism. A simple example is given by the joint distribution of pairs of neighbouring variables, setting $g$ to be the graph of two connected sites, we may define
\begin{equation*}
P(x,y):=P_g(x,y)=\frac{1}{|E|}\sum_{(i,j)\in E} \mathbb{I}\Big\{X_i=x,X_j=y\Big\}\,.
\end{equation*}
We will refer to distributions of the type (\ref{emp}), which are measured directly from an underlying stochastic process, as empirical. Note that empirical distributions will change with time, though we have not made this explicit in the notation. \par
The usual methodology of analyses based on objects of the type $P_g(\bx)$ is to move from regarding them as measurements of a stochastic system to treating them as dynamical variables in their own right. This is achieved by rescaling time by a factor of $N^{-1}$ and taking the limit $N\to\infty$ (the limit of large networks), in which it is assumed that $P_g(\bx)$ will approach a smooth non-random function, satisfying a differential equation which can be derived from the details of the underlying stochastic process. The complicating factor in this approach when applied to networked stochastic systems is that generally the rate of change of $P_g(\bx)$ will depend not only on those variables which contribute to $P_g(\bx)$, but also on their neighbours, leading to a system which is not closed. There is more than one way to perform this analysis and different sets of differential equations may be found (for example, compare the approaches to SIR epidemics taken in references \cite{Rozhnova2010} and \cite{Keeling1999}), however, a relatively general situation is described by a collection of differential equations of the form 
\begin{equation}
\frac{d}{dt}P_g(\bx_g)=\sum_{\by}P_G(\by)\Gamma(\bx_g,\by)\,, 
\label{gen_diff1}
\end{equation}
where $G$ is the graph containing those sites in $g$ and their neighbours, and the functions $\Gamma(\bx_g,\by)$ describe probability flows between states, derived from the microscopic dynamics. In this form, such equations are of very little use as they say that the rate of change of $P_g$ depends on the higher dimensional distribution $P_G$, which is not part of the system. Moment-closure techniques seek to address this problem by suggesting possible ways to express $P_G$ approximately as a function of $P_g$. In the next subsection, we will derive the maximum-entropy approach to moment-closure.
\subsection{Derivation}
Consider a graph $G$, covered by a collection $\mathbb{G}$ of subgraphs. To each site $i\in G$ a random variable $X_i$ is attached: write $\bm{X}$ for the vector containing these variables, and $\bm{X}_g$ for the projection of that vector onto the set of sites also included in the subgraph $g\in\mathbb{G}$. The $X_i$ are assumed to have a joint distribution function $P_G$ which is \emph{unknown}. Suppose that we have knowledge of the marginal distribution on each of the subgraphs, given by
\begin{equation*}
P_g(\bx_g):=\mathbb{P}\Big[\bm{X}_g=\bx_g\Big]=\sum_{x_i\,:\, i\notin g} P_G(\bx)\,.
\end{equation*}
In the context of moment-closure, the subgraph distributions $P_g$ are the level at which we wish to close our system of differential equations and $G$ is the larger graph containing all variables which influence the time evolution of a particular $P_g$. For a simple example, the pair approximation corresponds to the situation in which $G$ is a single site plus its neighbours, and $\mathbb{G}$ is the collection of edges in $G$.\par
The core of the problem is as follows: to close our system of differential equations at the level of the $P_g$, it is necessary to approximate the unknown higher-dimensional distribution $P_G$, guided only by knowledge of the marginal distributions $P_g$.\par 
It is correct to demand that the function $\tP_G$ approximating $P_G$ must of course be a probability distribution itself, and that its marginals $\tP_g$ should agree with the known distributions $P_g$, that is: 
\begin{equation}
\textrm{for all $\bx$, and for every $g\in\mathbb{G}$, we have }\qquad\sum_{x_i\,:\, i\notin g} \tP_G(\bx)=P_g(\bx_g)\,.
\label{marg_constraint}
\end{equation}
Aside from these requirements, in a general setting there is no \textit{a priori} reason to assume that $\tP_G$ should take any particular form. It can therefore be argued \cite{Jaynes1957} that the most sensible choice for $\tP_G$ is the distribution which has maximum entropy subject to the above constraint. Put simply, to make any other choice for $\tP_G$ would introduce an additional bias which would require further justification.\par
Accepting this reasoning, the calculation is straightforward. It will suffice to satisfy (\ref{marg_constraint}), as the requirement that $\tP_G$ is a bona fide distribution will then hold automatically. The entropy of a distribution $P$ is given by
\begin{equation*}
E(P)=-\sum_{\bx}P(\bx)\log P(\bx)\,.
\end{equation*}
To perform the constrained maximisation of this quantity, we introduce one Lagrange multiplier $\lambda_g(\bx_g)$ per subgraph $g\in\mathbb{G}$ and projected state vector $\bx_g$. We thus seek extrema of the functional
\begin{equation*}
\Lambda[P]=-\sum_{\bx}P(\bx)\log P(\bx)+\sum_g\sum_{\bx_g}\lambda_g(\bx_g)\left(P_g(\bx_g)-\sum_{x_i\,:\,i\notin g}P(\bx)\right)\,.
\end{equation*}
Differentiating, we find
\begin{equation*}
\frac{\partial \Lambda[P]}{\partial P(\bx)}=-1-\log P(\bx)-\sum_g\lambda_g(\bx_g)\,,
\end{equation*}
and thus the maximum entropy approximation for $P_G$ is given by
\begin{equation*}
\tP_G(\bx)=\exp\left\{-1-\sum_g \lambda_g(\bx_g)\right\}\,.
\end{equation*}
We see that $\tP_G$ factorises over subgraphs; there exist functions $Q_g$ such that
\begin{equation}
\tP_G(\bx)=\prod_g Q_g(\bx_g)\,,
\label{tP}
\end{equation}
where the equations for Lagrange multipliers imply that the functions $Q_g$ collectively satisfy
\begin{equation}
\textrm{for all $g$ and $\bx_g$, }\quad P_g(\bx_g)=\sum_{\by} \delta\big(\bx_g-\by_g\big)\prod_h Q_h(\by_h) \,.
\label{Lag_constraint}
\end{equation}
Note that the normalisation of the functions $P_g(\bx_g)$ was not accounted for during the implementation of the Lagrange multipliers, meaning that there are several redundant degrees of freedom in the solution set to (\ref{Lag_constraint}) which should be treated with care. Indeed, it is straightforward to see that (\ref{tP}) is invariant under certain multiplicative transforms of the $Q_g$, however, this is only a problem if one intends to work directly with the $Q_g$ in a numerical context.
\subsection{Practical implementation}
Although it is known \cite{Jaynes1957} that (provided the supplied marginals $P_g$ are consistent) the maximum entropy distribution $\tP_G$ always exists and is unique, only in certain circumstances can it be written neatly as a function of the marginals. We can characterise these situations by considering the graph $\mathcal{H}$, with sites given by the elements of $\mathbb{G}$ (that is, the subgraphs of $G$ for which marginals are supplied), and edge set given by the rule that $g\sim h$ if and only if $g$ and $h$ share at least one site. If $\mathcal{H}$ is a tree, then 
\begin{equation}
\tP_G(\bx)=\prod_g\frac{P_g(\bx_g)}{\sqrt{\prod_hP_{g\cap h}(\bx_{g\cap h})}}\,,
\label{trees}
\end{equation}
where $P_{g\cap h}$ is the marginal distribution of variables contained in both $g$ and $h$, and we set $P_\emptyset\equiv1$ for consistency.\par
For a general setting in which $\mathcal{H}$ is not a tree, there is no algebraic expression for $\tP_G$ in terms of the $P_g$. Numerical solution of the problem is straightforward however, for example by a simple process of iterative replacement we describe now. Starting from a uniform distribution $P^{(0)}$, one repeatedly cycles through the subgraphs $g\in\mathbb{G}$, at each step setting
\begin{equation*}
P^{(n+1)}(\bx)=P_g(\bx_g)\,\frac{P^{(n)}(\bx)}{\sum_{x_i\,:\,i\notin g}P^{(n)}(\bx)}\,.
\end{equation*}
As $n\to\infty$, the distributions $P^{(n)}$ converge to $\tP_G$. Proof of the convergence of this procedure may be found in \cite{Csiszar2004}, where it is employed in the entirely different context of quantifying distributional complexity by analysis of interaction structure. \par
The final consideration in the practical application of maximum-entropy moment-closure is the inclusion of the approximation in the differential equations it is intended to close. As discussed in the introduction, for a given microscopic dynamics there often exists more than one set of ODEs purporting to approximate the time evolution of the system. We consider the case that one wishes to close a system of differential equations of the form (\ref{gen_diff1})
which we assume are derived from some underlying stochastic process. In the fortunate situation that the graph $\mathcal{H}$ of subgraph overlaps is a tree, one may simply apply equation (\ref{trees}) to close the system by setting
\begin{eqnarray}
\frac{d}{dt}P_g(\bx_g)&=\sum_{\by}\Gamma(\bx_g,\by)P_G(\by)\nonumber\\
&\approx \sum_{\by}\Gamma(\bx_g,\by)\prod_{g'}\frac{P_{g'}(\by_{g'})}{\sqrt{\prod_hP_{{g'}\cap h}(\by_{g'\cap h})}}\,. 
\label{difftrees}
\end{eqnarray}
Away from this special case one only has access to $\tP_G$ via numerical methods, and no such simple expression is possible. However, if, as is very often the case, one intends to treat the differential equations numerically, then the lack of a simple closure formula is not much of a penalty. At each step of the numerical solution of the differential equations, the numerical value of $\tP_G$ may be recovered efficiently via the iterative replacement scheme described above. \par
Alternatively, the equations can always be closed at the level of the functions $Q_g$, with the resulting system given by, for all $g$ and $\bx_g$,
\begin{equation}
\sum_{\by}\left(\prod_hQ_h(\by_h)\right)\left[\Gamma(\bx_g,\by)-\delta(\bx_g-\by_g)\sum_h\frac{d}{dt}\log Q_h(\by_h)\right]=0\,.
\label{diffQ}
\end{equation}
As a result of the under-specification of the Lagrange multipliers discussed earlier, there are some extraneous degrees of freedom in this system of differential equations, though these can be accounted for either by holding constant a suitable number of values $Q_g(\bx_g)$, or by imposing additional normalisation constraints on the $Q_g$ to prevent divergence.\par
To summarise, in the general case for which equation (\ref{difftrees}) does not apply, one has the choice of working with $\tP_G$ numerically, or directly with the $Q_g$ via (\ref{diffQ}). 
\subsection{Example - pair mean field}
Consider the SIR model on a regular graph of degree $k$. In \cite{Keeling1999}, the progression of an SIR epidemic was modelled by differential equations equivalent to 
\begin{eqnarray}
&\frac{d}{dt}P(S,S)=-2\,\frac{\beta}{k}P(S,S,I)\nonumber\\
&\frac{d}{dt}P(S,I)=\frac{\beta}{k}\Big(P(S,S,I)-P(I,S,I)-P(S,I)\Big)\nonumber\\
&\frac{d}{dt}P(S,R)=-\frac{\beta}{k}P(R,S,I)+\gamma P(S,I)\label{keeling_odes}\\
&\frac{d}{dt}P(I,I)=2\,\frac{\beta}{k}\Big(P(I,S,I)+P(S,I)\Big)-2\,\gamma P(I,I)\nonumber\\
&\frac{d}{dt}P(I,R)=\frac{\beta}{k}P(R,S,I)+\gamma\Big(P(I,I)-P(I,R)\Big)\,.\nonumber
\end{eqnarray}
In these expressions $k$ denotes the average degree (number of neighbours) of the sites in the graph, $P(x,y)$ is probability distribution of a pair of adjacent variables, and 
\begin{equation}
P(x,y,z)=(1-\phi)P_R(x,y,z)+\phi P_T(x,y,z)\,,
\label{row_tri}
\end{equation}
where $P_R(x,y,z)$ is the distribution of three sites in a row, $P_T(x,y,z)$ is the distribution of three sites forming a triangle, and $\phi$ is a measure of the frequency of triangles in the underlying graph. To close these equations at the level of $P(x,y)$, we are required to find expressions for $P_R(x,y,z)$ and $P_T(x,y,z)$. Note that although these equations involve distributions of dimensions two and three and thus are not of the same form as (\ref{gen_diff1}), maximum-entropy moment-closure may still be applied. \par
\begin{figure}
\begin{center}
\includegraphics[width=0.5\textwidth]{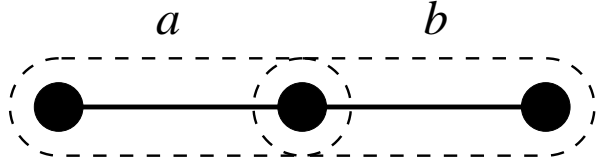}
\end{center}
\caption{The graph $R$ consisting of three sites arranged in a row, covered by subgraphs $a$ and $b$ given by linked pairs of sites.}
\label{row}
\end{figure}
Considering $P_R(x,y,z)$ first, we are interested in the situation illustrated in Figure \ref{row} -- we wish to compute the maximum-entropy approximation $\tP_R$, under the constraint that each of the pairs defined by subgraphs $a$ and $b$ has the known marginal distribution $P(x,y)$. From equation (\ref{tP}), we see that there exist functions $Q_a$ and $Q_b$ such that 
\begin{equation*}
\tP_R(x,y,z)=Q_a(x,y)Q_b(y,z)\,.
\end{equation*}
The constraints (\ref{Lag_constraint}) then give first for subgraph $a$,
\begin{eqnarray*}
& P(x,y)=P_a(x,y)=\sum_z Q_a(x,y)Q_b(y,z)=Q_a(x,y)\sum_zQ_b(y,z)\\
& \quad\Rightarrow\quad Q_a(x,y)=\frac{P(x,y)}{\sum_zQ_b(y,z)}\,,
\end{eqnarray*}
and then for subgraph $b$,
\begin{eqnarray*}
& P(y,z)=P_b(y,z)=\sum_x Q_a(x,y)Q_b(y,z)=Q_b(y,z)\sum_x \frac{P(x,y)}{\sum_zQ_b(y,z)}\\
& \quad\Rightarrow\quad Q_b(y,z)=\frac{P(y,z)}{P(y)}\sum_zQ_b(y,z)\,.
\end{eqnarray*}
Combining the two yields
\begin{equation}
\tP_R(x,y,z)=\frac{P(x,y)P(y,z)}{P(y)}\,.
\label{pair_row}
\end{equation}
Alternatively, we might observe that there are only two subgraphs and we may then apply the simple form (\ref{trees}) directly to obtain the same result.\par
Equation (\ref{pair_row}) is precisely the usual pair approximation to the joint distribution of three variables connected in a row, so we see that the maximum-entropy approach agrees with the usual methods in this case. 
\begin{figure}
\begin{center}
\includegraphics[width=0.3\textwidth]{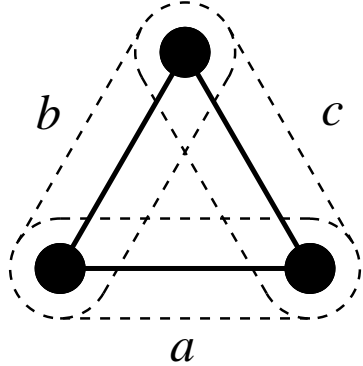}
\end{center}
\caption{The graph $T$ consisting of three sites arranged in a triangle, covered by subgraphs $a$, $b$ and $c$, given by linked pairs of sites.}
\label{tri}
\end{figure}
Turning attention now to the case of the triangle (illustrated in Figure \ref{tri}), we find that the situation is not so straightforward. The graph of subgraph overlaps in this case is not a tree and thus there is no simple algebraic form for $\tP_T$. In \cite{Matsuda2000, Keeling1997} an approximation is suggested equivalent to 
\begin{equation}
P_T(x,y,z)\approx F(x,y,z):=\frac{P(x,y)P(y,z)P(z,x)}{P(x)P(y)P(z)}\,.
\label{triK}
\end{equation}
This is expression is sometimes \cite{Matsuda2000,Filipe2003} attributed to Kirkwood \cite{Kirkwood1942}. Unfortunately, for almost all $P_T$ this function is not a distribution itself and does not agree with the known marginals on pairs. In defense of this expression, it is usually clearly advertised as an approximation made for convenience, and besides we will see later that it holds up remarkably well when used to numerically integrate equations (\ref{keeling_odes}), despite being inaccurate in general. Note that (\ref{triK}) is precisely the result one would obtain from naively (incorrectly) applying (\ref{trees}) in this situation. \par
For numerical applications, the maximum-entropy approximation to $P_T$ may easily be computed using the iterative replacement algorithm outlined earlier. Such an approach may however be unappealing to those who would like a neat closed system of ODEs, whose terms do not depend on quantities which must be computed numerically. In that case, one might choose to halt the iterative replacement process after some number of steps. For example, three steps will yield the approximate form
\begin{equation}
P_T(x,y,z)\approx \widehat{P}_T(x,y,z)=\frac{\displaystyle P(x,y)P(y,z)P(z,x)}{\displaystyle\sum_w P(x,w)P(z,w)\frac{P(x)}{P(w)}}\,,
\label{triT}
\end{equation}
which is a viable alternative to (\ref{triK}) with the advantages firstly of being a genuine probability distribution and secondly of agreeing with the marginal distribution on the pair $(x,y)$. As will be reported in the next section, numerical experiments comparing the approximations (\ref{triK}) and (\ref{triT}) have surprising results.
\section{Numerical experiments}
\subsection{An epidemic on the square lattice}
As a first test of the utility of the maximum-entropy approach, we attempt to predict the behaviour of the SIR model on a square lattice by numerically integrating a moment-closed system of equations. The pair approximation on a square lattice has already been considered in \cite{DeSouza2010}, so we will develop a higher order theory based on clusters of five connected sites.\par
\begin{figure}
\centering
\subfloat[The graph $g$\label{g}]{\includegraphics[width=0.17\textwidth, trim=0 -75 0 0]{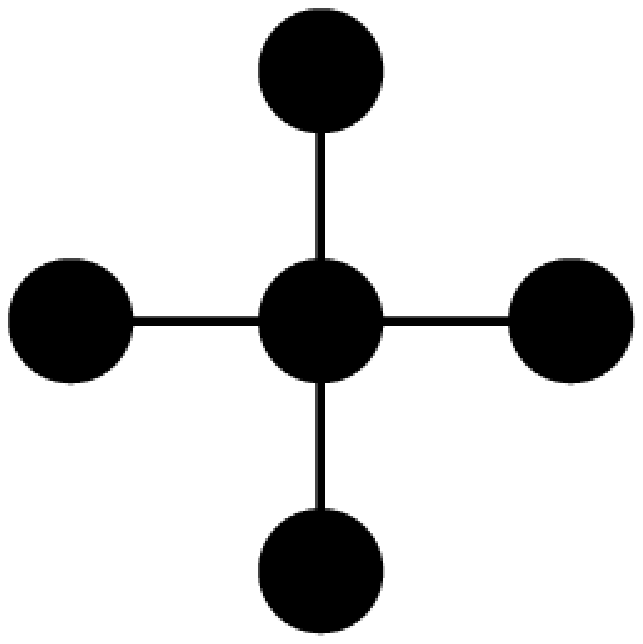}}\qquad
\subfloat[The graph $G$ of $g$ and its neighbours in the lattice\label{G}]{\includegraphics[width=0.3\textwidth, trim=-18 -20 -18 0]{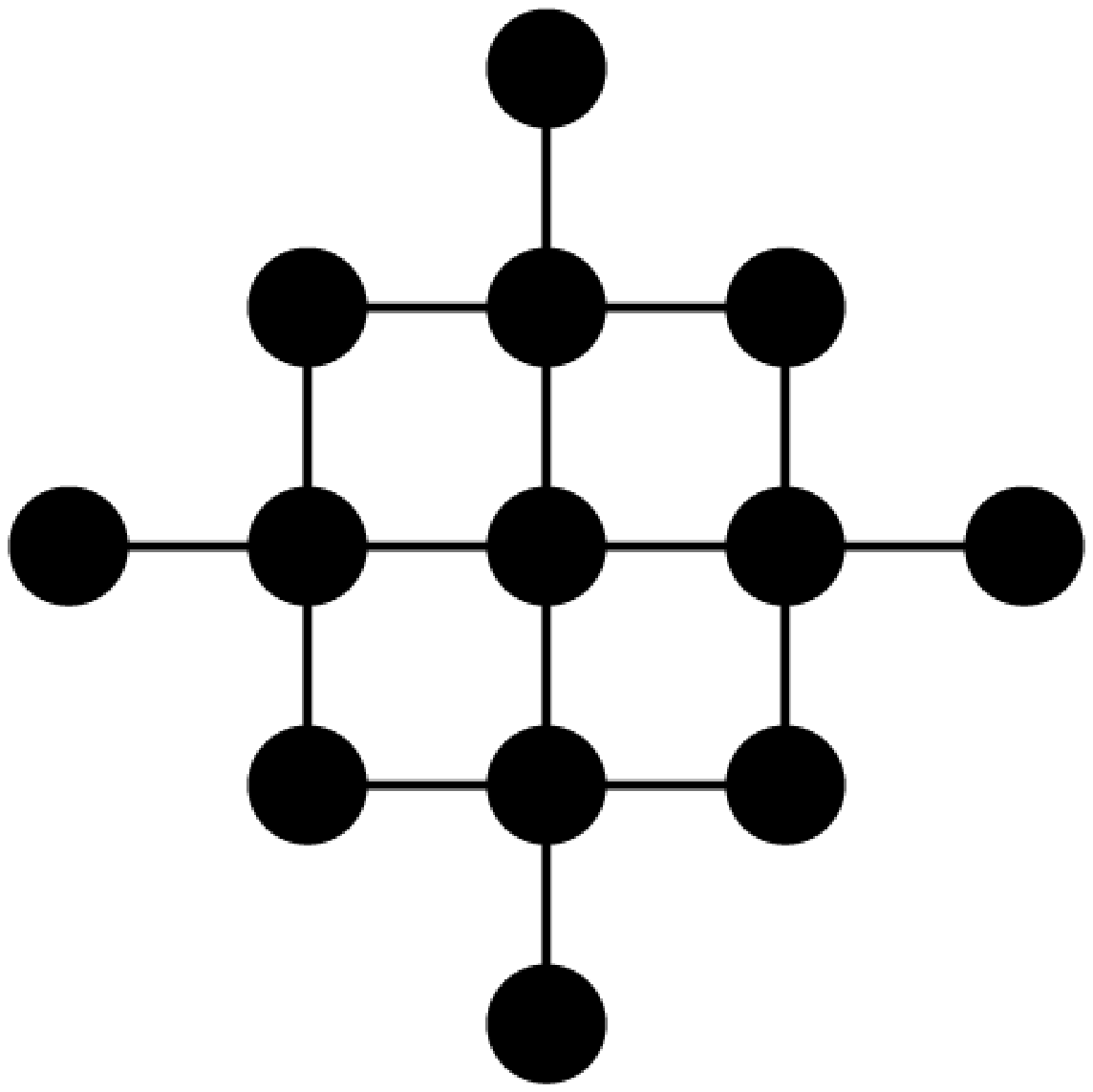}}\qquad
\subfloat[The collection $\mathbb{G}$ of five copies of $g$ which cover $G$\label{GG}]{\includegraphics[width=0.3\textwidth]{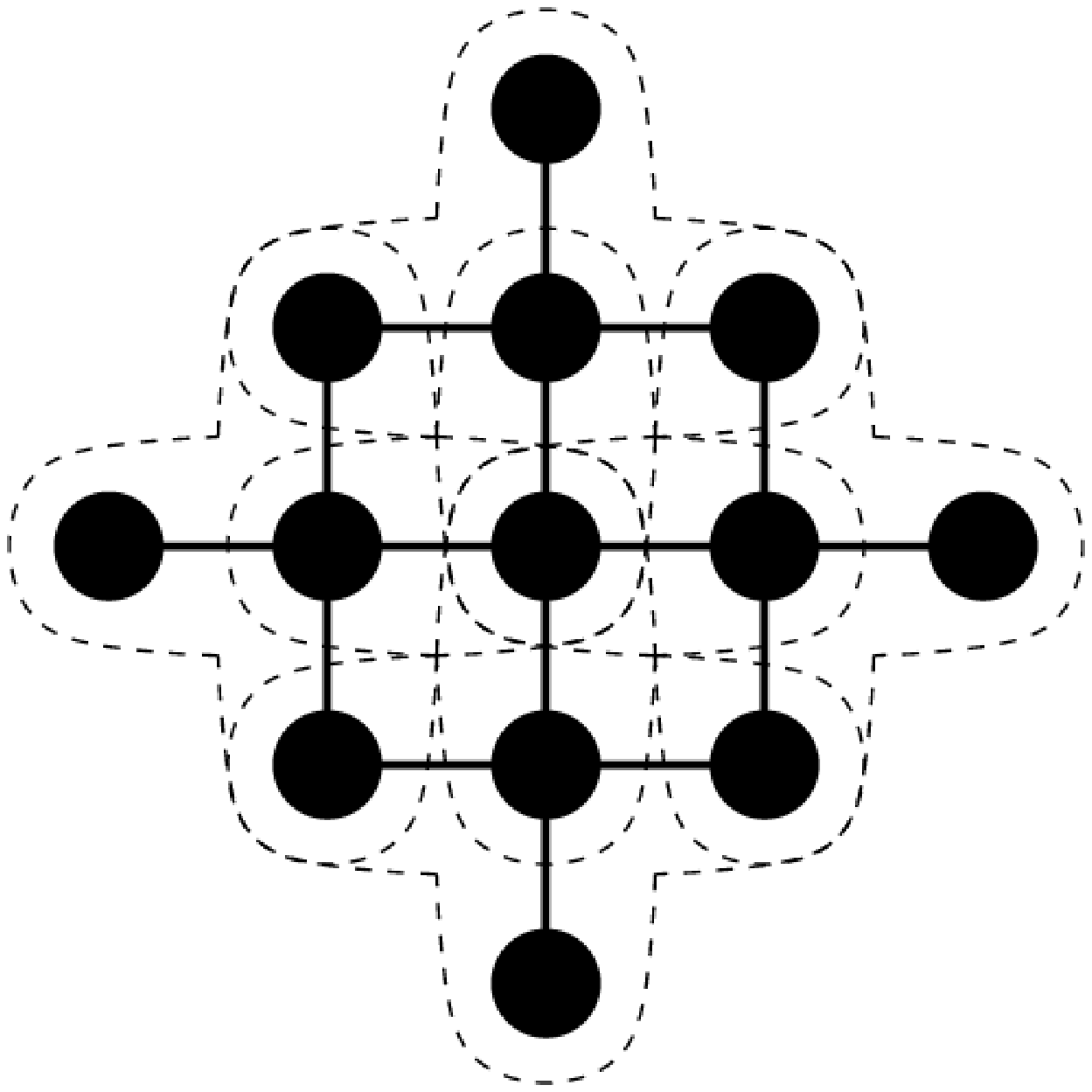}}
\caption{Developing a theory for the SIR model on a square lattice which is closed at the level of $P_g$ (a site and its neighbours) requires estimating the distribution of variables in $P_G$, which itself contains five copies of $g$.}
\label{lat}
\end{figure}
Let $g$ be the graph of a single site and its neighbours in the square lattice -- see Figure \ref{g} for an illustration. To close the equations for SIR epidemics at the level of $P_g$ will require approximation of the distribution $P_G$ on the graph $G$ of all sites within distance two of a central site (Figure \ref{G}). This bigger graph contains five copies of $g$ (Figure \ref{GG}), which are used as the basis of the maximum-entropy approximation. \par
As per the earlier discussion, one formulation for the moment-closed system of differential equations is given by
\begin{equation}
\frac{d}{dt}P_g(\bx_g)=\sum_{\by}\widetilde{P}_G(\by)\Gamma(\bx_g,\by)\,, 
\label{gen_diff}
\end{equation}
however, to compute the coefficients $\Gamma(\bx_g,\by)$ one must enumerate the possible changes to each $h\in\mathbb{G}$ which could occur given the configuration $\by$ of the sites in $G$. Explicitly,
\begin{equation*}
\Gamma(\bx_g,\by)=\sum_{h\in\mathbb{G}}\left[\mathbb{P}(\by_h\mapsto\bx_g\,|\,\by)-\mathbb{I}\Big\{\bx_g=\by_h\Big\}\Big(1-\mathbb{P}(\bx_h\mapsto\bx_h\,|\,\by)\Big)\right]\,,
\end{equation*}
where $\mapsto$ is used as a shorthand for a change taking place in one timestep. \par
Note that for SIR dynamics there are three possible states per site which, given that there are five sites in $g$ and thirteen in $G$, means that the system (\ref{gen_diff}) contains $243$ equations and moreover each one has $1594323$ terms in its sum. The solving of this system is most certainly a job for the computer - see \cite{Alexander2010} for a discussion of this problem in relation to the pair approximation.\par
\begin{figure}
\psfrag{Fraction Susceptible}{Fraction Susceptible}
\psfrag{Time}{Time}
\centering\includegraphics[width=\textwidth, trim=100 0 100 0]{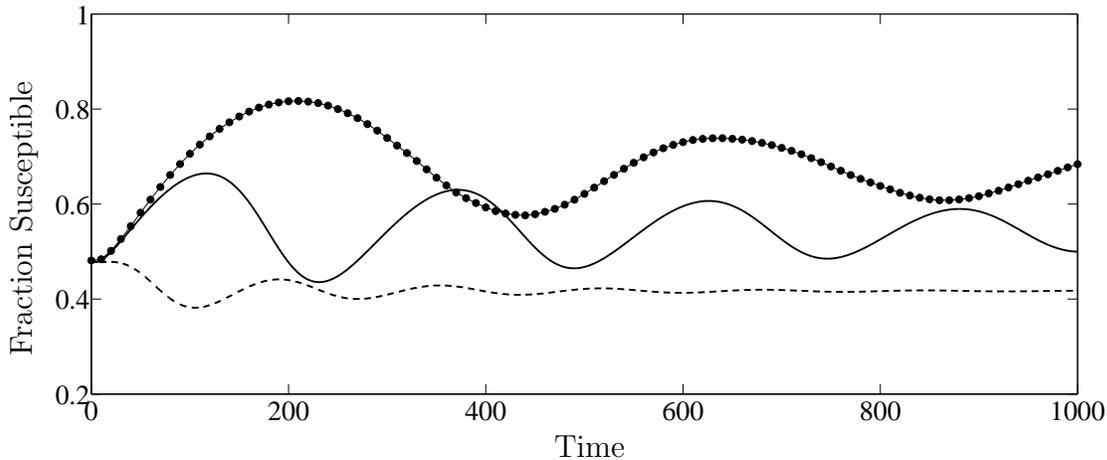}
\caption{Comparison of simulation results (circles) of SIR dynamics on a square lattice with predictions of the pair approximation (dashed line) and the five site maximum-entropy approximation (solid line) described in the text. The diesease parameters are $\beta=0.6$, $\gamma=0.1$ and $\alpha=0.01$.}
\label{latbad}
\end{figure}

Instead of attempting to list the equations themselves, a better understanding can be gained through an examination of the steps involved in numerically integrating them. Starting from an initial condition in which each state of each site is chosen independently at random,
\begin{equation*}
P_g(\bx_g)=\prod_{i\in g} P(x_i)\,,
\end{equation*}
we apply the Euler forward method of numerical integration. At each step, time is moved on by a small amount $t\mapsto t+\Delta t$, and the following computation is performed:
\begin{enumerate}
 \item Initially, for each $\bx_g$, set $$\frac{d}{dt}P_g(\bx_g)=0$$
 \item For every configuration $\by$ of variables in $G$
\begin{enumerate}
 \item Compute the maximum-entropy approximation $\widetilde{P}_G(\by)$ by iterative replacement
 \item Compute the probability $\delta$ of a change to the state of the central variable in this configuration, write $\by'$ for the resulting configuration
 \item For each $h\in G$ put $$\frac{d}{dt}P_g(\by_h)\mapsto \frac{d}{dt}P_g(\by_h)-\delta \widetilde{P}_G(\by)$$
 and $$\frac{d}{dt}P_g(\by'_h)\mapsto \frac{d}{dt}P_g(\by'_h)+\delta \widetilde{P}_G(\by)$$
\end{enumerate}
 \item For each $\bx_g$, update
 $$P_g(\bx_g)\mapsto P_g(\bx_g)+\Delta t\,\left( \frac{d}{dt}P_g(\bx_g)\right)\,.$$
\end{enumerate}
\begin{figure}
\psfrag{Fraction Infected}{Fraction Infected}
\psfrag{Time}{Time}
\centering\includegraphics[width=\textwidth, trim=100 0 100 0]{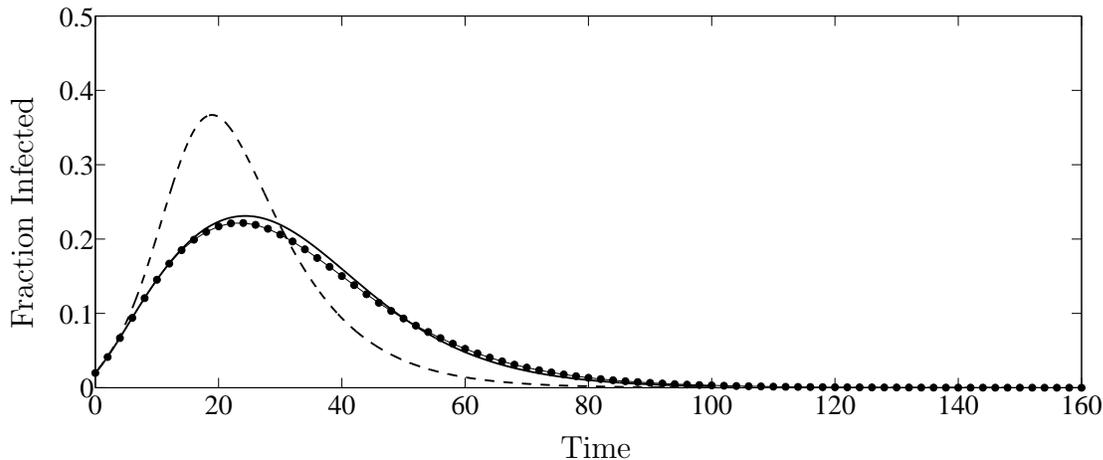}
\caption{Comparison of simulation results (circles) of SIR dynamics on a square lattice with predictions of the pair approximation (dashed line) and the five site maximum-entropy approximation (solid line) described in the text. The diesease parameters are $\beta=0.6$, $\gamma=0.1$ and $\alpha=0$.}
\label{latgood}
\end{figure}
The results of this process for two different parameter choices are presented in Figures \ref{latgood} and \ref{latbad}. In each case a single random simulation of a discrete time SIR epidemic on a toroidal square lattice of size $N=10^6$ was performed according to the microscopic dynamics outlined in Section 2.1.\par 
For Figure \ref{latgood} the parameters for infection, recovery and return to susceptibility were chosen to be $\beta=0.6$, $\gamma=0.1$ and $\alpha=0$, respectively. Alongside the result from the simulation are those obtained by the numerical integration scheme for the cluster approximation described above, as well as the pair approximation of \cite{DeSouza2010} for comparison. As is clearly visible in the figure, the cluster approximation provides a considerable improvement to the accuracy of the prediction. \par
An identical set-up was used for Figure \ref{latbad}, with the exception that return to susceptibility was allowed with probability $\alpha=0.01$. This change has the effect of moving the system into an oscillatory regime in which both the pair approximation and the cluster approximation are dramatically less accurate. The failure of moment-closure methods in this regime can be attributed to the development of long range correlations in the model which appear to play a central role in controlling the frequency of oscillations observed. It is worth noting, though, that for many practical applications the goal of moment-closure methods is to quantify the steady states of the system (and their stability), rather than to accurately reproduce transient features such as the decaying oscillations observed here. In this regard, the cluster approximation is again a clear improvement over the pair approximation. 
\subsection{Comparison of pair approximations}
In the previous section, we saw that the maximum-entropy approximation for three sites in a row agreed with the usual pair approximation for such a configuration. For a triangular arrangement of sites, however, the situation is rather more complicated. The commonly cited approximation $F$, given in equation (\ref{triK}) does not satisfy the requirement of being a probability distribution. An alternative expression $\widehat{P}_T$ was suggested here in equation (\ref{triT}), found as the partial implementation of maximum-entropy, and may possibly offer an improvement. \par
The following pair of numerical experiments are designed to give a casual assessment of the performance of these approximations. Firstly, the outputs are compared directly to the empirical distribution they are intended to approximate, secondly, the expressions $F$ and $\widehat{P}_T$ are employed in the numerical integration of the differential equations (\ref{keeling_odes}) to see how success or failure in the first test may affect the predictive power of the system they close.\par
The approximations in question deal with triangles, so a network with plenty of these will provide an appropriate setting for the numerical tests. We choose a small-world network constructed as follows: $N$ sites are arranged in a circle and each is connected to its four nearest neighbours, each edge then has one end randomly rewired with probability $\delta$ ($=0.1$ for the present case). Random graphs of this type have frequently been used as models of social networks and thus the spread of disease on such networks is interesting in its own right.  For the simulations used here a single graph of this type of size $N=1000$ was generated in this way. The frequency of triangles in this graph was measured at $\phi=0.355$. A sample of 1000 independent runs of SIR dynamics were then performed on this graph, using an initial condition of 20 infective sites chosen at random, and parameters for infection, recovery and return to susceptibility given by $\beta=0.8$, $\gamma=0.2$ and $\alpha=0$, respectively.  \par
\begin{figure}
\psfrag{Fraction Infected}{Fraction Infected}
\psfrag{Time}{Time}
\subfloat[Top -- empirical distribution $P_T(\,\cdot\,|S,I)$, computed as described in the text,\newline Centre -- prediction of the same quantity using equation (\ref{triT}),\newline Bottom -- prediction of the same quantity using equation (\ref{triK}).]{\includegraphics[width=0.8\textwidth]{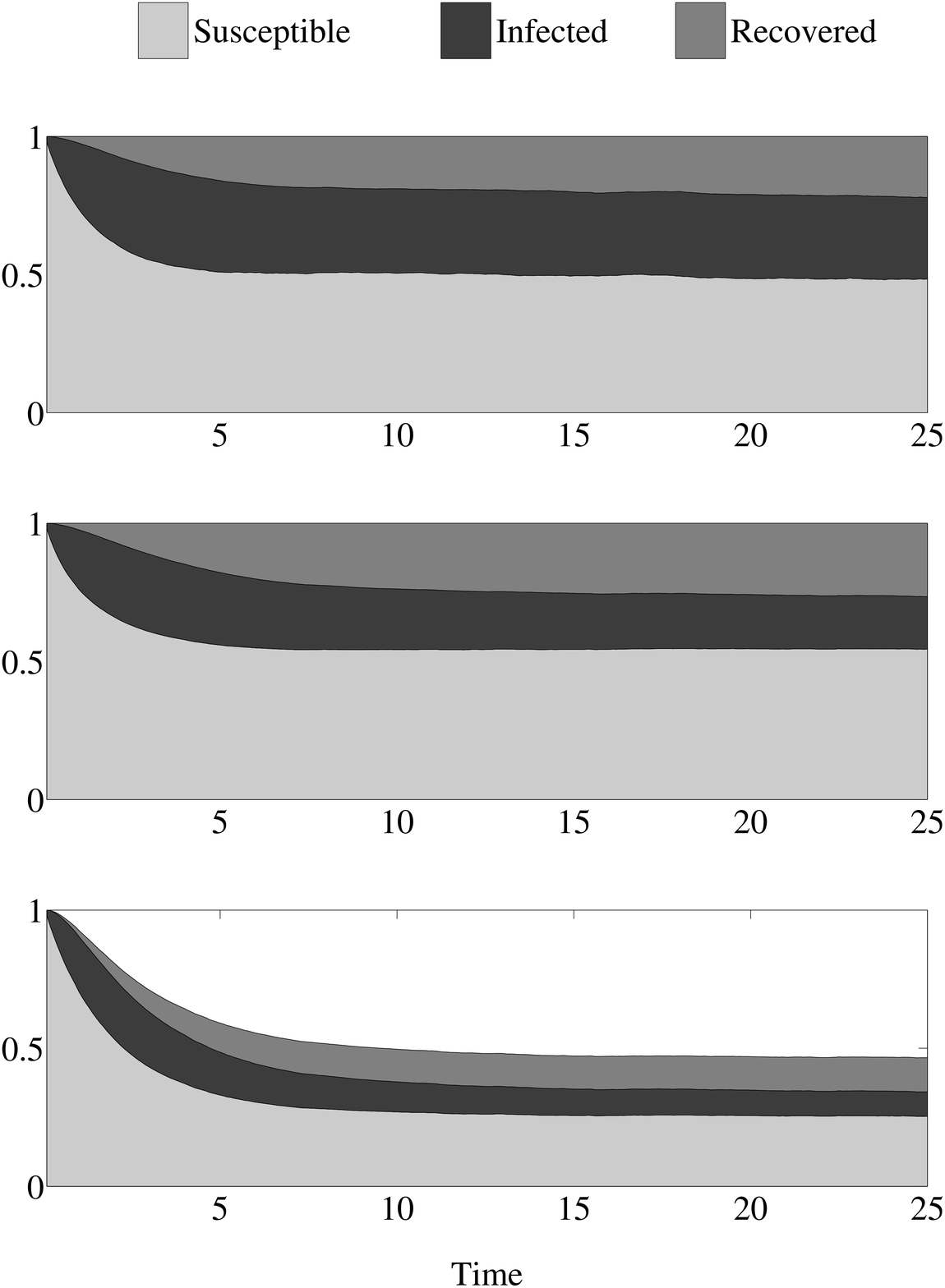}\label{pmf_comp_P}}\\
\centering \subfloat[Circles -- fraction of infected sites on a small-world network as described in the text,\newline Solid line -- prediction of the same quantity using approximation (\ref{triT}) in system (\ref{keeling_odes}),\newline Dashed line -- prediction of the same quantity using approximation (\ref{triK}) in system (\ref{keeling_odes}).]{\includegraphics[width=\textwidth, trim=160 0 0 0]{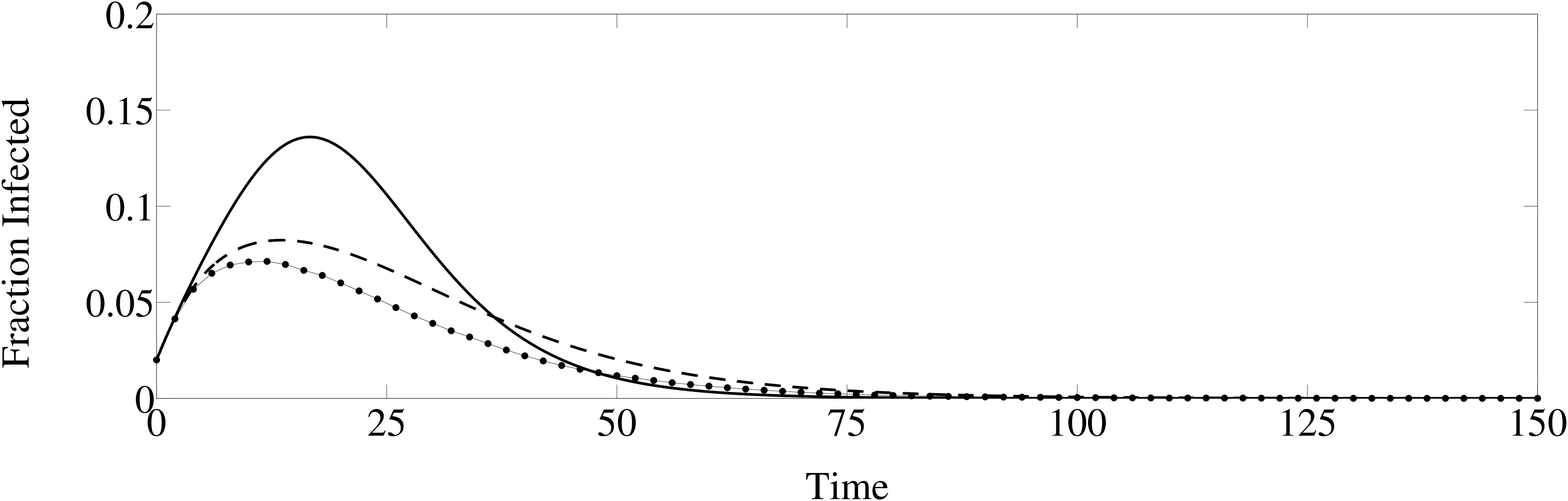}\label{pmf_comp_Inf}}
\caption{}
\end{figure}
For the first test, at each tenth of a timestep (this is possible since time is rescaled by a factor if $N^{-1}$ in the microscopic dynamics) the empirical distribution $P_T(x,y,z)$ of three sites arranged in a triangle was computed using equation (\ref{emp}). The empirical marginal distribution $P(x,y)$ of pairs of edges \emph{which are involved in a triangle} was computed by summing $P_T(x,y,z)$ over one argument. \par
From equations (\ref{keeling_odes}) and (\ref{row_tri}), one sees that the quantities in need of approximation are $P_T(S,S,I)$, $P_T(I,S,I)$ and $P_T(R,S,I)$, which between them describe the distribution of the state of the third site in a triangle in which the others form a susceptible-infected pair. The shaded areas in the top plot of Figure \ref{pmf_comp_P} shows the empirical conditional density $P_T(\,x\,|S,I)=P_T(x,S,I)/P(S,I)$ for $x=S,I,R$, obtained from the simulation. Since they form a probability distribution, they necessarily sum to one. For the central plot of Figure \ref{pmf_comp_P}, the approximate conditional density $\widehat{P}_T(\,x\,|S,I)/P(S,I)$ at each data point was computed using equation (\ref{triT}), with the empirical pair distribution $P(x,y)$ taken as the input. Finally, the lower plot of the same figure shows the result $F$ of equation (\ref{triK}), computed in the same way. \par
The difference in the accuracy of the approximations $F$ and $\widehat{P}_T$ shown in Figure \ref{pmf_comp_P} is striking. Whilst $\widehat{P}_T$ appears to give a reasonably close approximation to the empirical conditional distribution $P_T(\,\cdot\,|S,I)$, the results of $F$ are a dramatic underestimate, particularly of the fraction of susceptibles. \par

In the production of Figure \ref{pmf_comp_P}, the empirical pair distribution $P(x,y)$ was used for as the basis for $F$ and $\widehat{P}_T$. If instead the approximations are employed for their intended purpose of closing the system of equations (\ref{keeling_odes}), the pair distribution used will evolve over time according to these equations and, as a result, errors may quickly become compounded to produce an inaccurate prediction of the progress of the disease. In Figure \ref{pmf_comp_Inf}, the time evolution of the fraction of infected sites as averaged over the simulations (circles) is shown alongside the predictions resulting from system (\ref{keeling_odes}) closed using approximations $F$ (dashed line) and $\widehat{P}_T$ (solid line). \par
As is clearly visible in the figure, the success of equation $\widehat{P}_T$ as an approximation to the distribution $P_T$ has not translated into a good prediction when used to close the system of differential equations. In contrast, the approximation $F$ is relatively successful when used in this way, providing a much closer fit to the data from numerical simulations.\par 
That the dramatic failure of $F$ in reconstructing the density $P_T$ does not result in a poor prediction of the progress of the disease is surprising and requires explanation. Note that both approximation schemes make use of the same estimate $\widetilde{P}_R$ for the distribution $P_R$ of three sites in a row, which here accounts for around 65\% of the estimate of $P(x,y,z)$, see equation (\ref{row_tri}). Performing a similar numerical experiment to assess the accuracy of this shared component reveals that (for this graph and parameter choice at least) it provides a substantial overestimate of the conditional density of susceptibles $P_R(S\,|S,I)$. In contrast, Figure \ref{pmf_comp_P} shows that approximation $F$ gives a large underestimate of the quantity $P_T(S\,|S,I)$, which has the effect of lowering the overall estimate of $P(S,S,I)$, acting to cancel out most of the erroneous overestimation introduced by $\widetilde{P}_R$. Since the approximation $\widehat{P}_T$ provides a much better estimation of $P_T(S\,|S,I)$, the error in the shared component $\widetilde{P}_R$ is carried over unchecked, resulting in a poorer overall performance.\par 
To summarise, it appears that the success of the approximation scheme proposed in \cite{Matsuda2000,Keeling1999} may in fact be largely due to a fortunate cancellation caused by two poor approximations giving large errors in opposite directions. As the experiments of this subsection have demonstrated, improving one part of the scheme actually has the counter productive effect of destroying this curious symmetry of errors, resulting in a worse performance overall. 
\subsection{Sensitivity to errors in approximation}
The central premise of moment-closure methods, including the maximum-entropy approach discussed in this article, is that having access to a `good' approximation scheme to higher order correlations will translate into a high accuracy of prediction from a system of differential equations employing them. It is therefore sensible to ask: precisely how good is `good'? To put this query in concrete terms, the following numerical experiment tests the sensitivity of moment-closure to errors in the approximation.\par
Consider the following system of differential equations describing the time evolution of the fraction of susceptible, infected and recovered individuals in a standard SIR epidemic:
\begin{eqnarray}
\frac{d}{dt}P(S)=-\beta P(S) P(I|S)+\alpha P(R)\nonumber\\
\frac{d}{dt}P(S)=\beta P(S) P(I|S)-\gamma P(I)\label{single_odes}\\
\frac{d}{dt}P(S)=\gamma P(I)-\alpha P(R)\,.\nonumber
\end{eqnarray}
Here the term $P(I|S)$ denotes the conditional probability that a randomly chosen neighbour of a randomly chosen susceptible site is infected, so for example $\beta P(S)P(I|S)$ gives the probability that a susceptible site is chosen for update and is then infected by a neighbour. These equations give an exact description (on average and in the limit $N\to\infty$) of the progress of an epidemic, however they are not closed as they depend on the unknown quantity $P(I|S)$. \par
Now consider the following numerical experiment. SIR dynamics are performed on a large square lattice and at each micro-timestep the empirical conditional probability $P(I|S)$ is measured directly from the simulation. Simultaneously with this process, the system of differential equations (\ref{single_odes}) is numerically integrated using the numerical value of $P(I|S)$ taken from the empirical measurement at the same moment in time. Since the equations  (\ref{single_odes}) are exact and $P(I|S)$ is determined from the simulation, the output from the simulation and the numerical integration will be almost indistinguishable -- this fact is easily verified on the computer. \par
By repeating this process, but now replacing $P(I|S)$ in the numerical integration by a slightly perturbed value, one will gain some insight into the robustness of the system (\ref{single_odes}) to errors in approximation of $P(I|S)$.\par
\begin{figure}
\psfrag{Fraction Recovered}{Fraction Recovered}
\psfrag{Time}{Time}
\centering\includegraphics[width=\textwidth, trim=100 100 100 0]{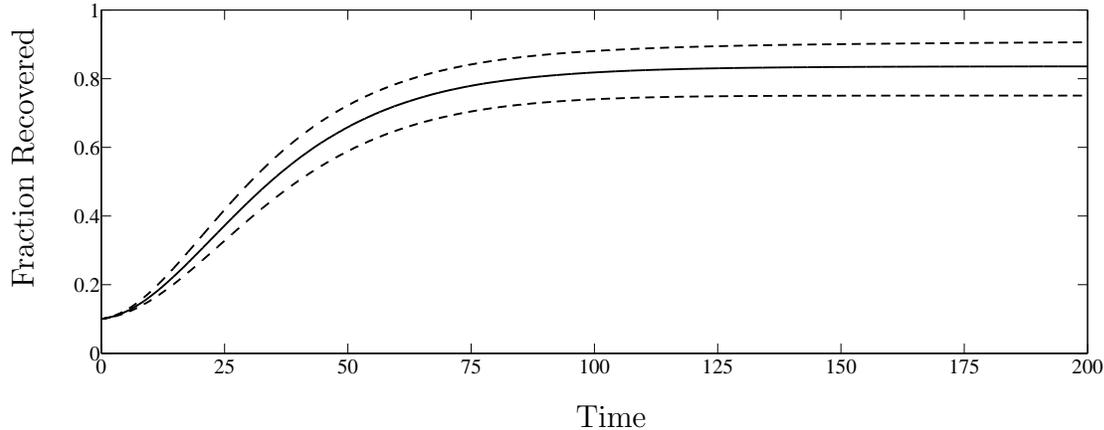}
\caption{Solid line -- fraction of recovered sites measured from a single run of SIR dynamics on a toroidal square lattice of size $N=10^6$. Dashed lines -- high and low results from numerical integration of differential equation system (\ref{single_odes}), with the pair distribution modified but within $\varepsilon=0.001$ in KL-divergence from the empirical value.}
\label{error}
\end{figure}

The results of such an experiment are shown in Figure \ref{error}. A single run of SIR dynamics was performed on a toroidal square lattice of size $N=10^6$, with parameters $\beta=0.6$, $\gamma=0.1$ and $\alpha=0$. The high and low perturbations of $P(I|S)$ were obtained by adjusting this value up or down as far as possible whilst keeping the Kullback Leibler divergence\footnote{KL-divergence is a measure (not a metric) of the difference between two probability distributions, defined in the discrete case by $D(P_1\|P_2)=\sum_x P_1(x)\log [P_1(x)/P_2(x)]$.} from the empirical distribution $P(\,\cdot\,|S)$ less than $\varepsilon=0.001$.\par
For most practical purposes, a KL-divergence of $\varepsilon=0.001$ is extremely low and would represent a very close approximation to the true distribution. As demonstrated in Figure \ref{error}, however, a such small range of error in the distribution can still result in considerable uncertainty in the resulting prediction of system (\ref{single_odes}), as the error is quickly compounded over the course of the epidemic. Even with such a precise approximation to $P(I|S)$, the final size of the epidemic may be over- or under-estimated by around 10\%. \par
On first glance it may seem unfair to choose extremal errors in $P(I|S)$, however, it should be noted that, as we saw in the previous subsection, errors in moment-closure approximations may often be systematic rather than random and hence such a choice is quite realistic.
\section{Conclusion}
Moment-closure methods are vital tools in the analysis of stochastic systems with network structure, allowing one to make mathematical predictions about the behvahiour of models previously only accessible to simulation. The success of such methods is certainly not in question, however, it is mitigated by several issues:
\begin{enumerate}
\item the most common moment-closure method (the pair approximation) is inconsistent in its treatment of triangles
\item higher order approximation schemes usually involve an arbitrary and unmotivated choice of some conditional independence structure
\item moment-closed systems of differential equations (particularly those of higher order) are computationally intensive to solve
\item the insertion of an approximation scheme into a system of differential equations obfuscates the effects of the accuracy of the approximation.
\end{enumerate}
The work presented in this article provides an effective solution to the first two issues on this list. By following a systematic approach based on maximisation of entropy, one is able to obtain approximation schemes of any desired size and complexity, without the need for any arbitrary or inconsistent assumptions.\par
In those situations in which a conditional independence structure genuinely is present, such as the pair approximation of three sites in a row, it is recovered by maximum-entropy and simple algebraic equations for the approximation may be derived. In more general situations where no such simple equations exist, the maximum-entropy approximation may be computed efficiently by a process of iterative replacement. Resorting to numerical approaches is certainly an aesthetic failing but it is not a practical one as moment-closed systems of differential equations are already computationally intensive to solve and the use of the maximum-entropy approximation causes very little additional burden. Indeed, numerical computation of the closure terms is all that is required for analysis using software packages such as AUTO \cite{Doedel1997}, and so the maximum-entropy method may be straightforwardly implemented there. \par
To investigate the application of maximum-entropy moment-closure, issue (ii) above was considered in some detail. The usual pair approximation to triangles suffers from the mathematical drawbacks of neither being properly normalised, nor agreeing with the known marginal distribution on pairs. The maximum-entropy approximation by definition does not suffer either of these problems, however, it cannot be written down in a neat closed form. As a middle ground, an alternative pair approximation to triangles was obtained by halting the iterative replacement algorithm after three steps. In numerical experiments based on the SIR model on a small world network, this alternative scheme vastly outperformed the usual approximation in the job of predicting the joint distribution of three sites forming a triangle. However, when coupled with the approximation for three sites in a row and used to close a system of differential equations, the results of the new approximation did not agree with those of numerical simulations, whilst making the less accurate approximation to triangles somehow resulted in a better fit to the data. Further numerical investigation revealed that this unusual effect is due to a fortuitous cancellation of error between the pair approximations for rows and triangles, whose balance was upset by improving one half of the approximation scheme but not the other.\par
The phenomenon that an improved approximation scheme can lead to worse results when used to close a system of differential equations raises often overlooked questions about the reliability and sensitivity of moment-closure methods in general. As the simple numerical experiment presented in Section 3.3 shows, an approximation which is highly accurate but suffers from a very small systematic bias can yield results with errors in prediction which are several orders of magnitude larger. This suggests that predictions made by moment-closure methods (whose biases we do not know) should be viewed as being somewhat flexible.\par
Moving forward, there is considerable scope to apply the methods outlined in this article to the study of stochastic systems on networks. It would be particularly interesting to take a maximum-entropy approach to the study of moment closure methods in adaptive networks \cite{Gross2006,Gross2008,Zschaler2010,Demirel2011} and stochastic corrections \cite{Rozhnova2010}.
\section*{Acknowledgments}
Thanks to Alan McKane, Tobias Galla and Joe Challenger for useful discussions and advice. Funding from the EPSRC under grant number EP/H02171X/1 is gratefully acknowledged.
\newpage
\bibliographystyle{iopart-num}
\bibliography{CMF}
\end{document}